\documentclass[preprint,12pt]{elsarticle}

\usepackage{amssymb}
\usepackage{amsmath}
\usepackage{xcolor}
 \usepackage{lineno}
 \usepackage{algorithm}
 \usepackage[noend]{algpseudocode}
 \usepackage{subcaption}

\journal{JQSRT}

\begin{document}

\begin{frontmatter}



\title{A modified recursive transfer matrix algorithm for radiation and scattering computation of a multilayered sphere}


\author[label1,label2]{Jianing Zhang}
%
\affiliation[label1]{organization={National Frontiers Science Center for Industrial Intelligence and Systems Optimization,  Northeastern University},
            city={Shenyang},
            country={China}}
 \affiliation[label2]{organization={College of Information Science and Engineering, Northeastern University},
            city={Shenyang},
            country={China}}

\begin{abstract}  
We discuss the electromagnetic scattering and radiation problems of multilayered spheres, reviewing the history of the Lorentz-Mie theory and the numerical stability issues encountered in handling multilayered spheres. By combining recursive methods with the transfer matrix method, we propose a modified transfer matrix algorithm designed for the stable and efficient calculation of electromagnetic scattering coefficients of multilayered spheres. The new algorithm simplifies the recursive formulas by introducing Debye potentials and logarithmic derivatives, effectively avoiding numerical overflow issues associated with Bessel functions under large complex variables. Moreover, by adopting a hybrid recursive strategy, this algorithm can resolve the singularity problem associated with logarithmic derivatives in previous algorithms. Numerical test results demonstrate that this algorithm offers superior stability and applicability when dealing with complex cases such as thin shells and strongly absorbing media.
\end{abstract}




\end{frontmatter}



\section{Introduction}

The problem of electromagnetic wave scattering by spherical particles was first proposed and addressed by Mie, Lorentz, and Debye, and is commonly known as the Lorentz-Mie theory or the Lorentz-Mie-Debye theory. For over a century, this theory has been an important topic of research in various fields such as electromagnetism \cite{chew95}, optics \cite{bohren83},  thermal and statistical physics \cite{chen08, kruger12}. Numerous monographs and papers have discussed this problem. On the other hand, the thermal fluctuations of electromagnetic waves are a core issue in radiative transfer, explaining groundbreaking discoveries such as Planck's law of blackbody radiation, proposed nearly a century ago. Electromagnetic thermal radiation itself can also be concisely expressed using the scattering operator of an object. The study of electromagnetic scattering and radiation of (multilayered) spheres is of fundamental importance for research in physical theory and has practical significance in applications such as remote sensing and imaging, biomedicine, photovoltaic design, and optical antennas.

As early as 1951, Aden and Kerker \cite{aden51} extended the Lorentz-Mie theory to the case of coated spheres. For multilayered spheres, the electric and magnetic fields are represented as multipole expansions, with the coefficients of these expansions determined by the boundary conditions at the interfaces of the sphere's layers. Kerker \cite{kerker57} further extended this theory to multilayered spheres. In the early days, applying Kerker's formulas to calculate the scattering of multilayered spheres often encountered various numerical issues, partly due to the use of incorrect recurrence formulas for calculating Bessel functions. Even with correct recurrence formulas, numerical problems still arose when calculating thin-shell spheres, strongly absorbing media, and highly multilayered cases. To address these numerical stability issues, Toon and Ackerman \cite{toon81} reformulated Kerker's equations using logarithmic derivatives, providing a stable algorithm for scattering by double-layered spheres. Further, Wu and Wang \cite{wu91} proposed a concise recursive method to calculate electromagnetic scattering coefficients for arbitrary multilayered spheres. Johnson \cite{johnson96} also developed a slightly different recurrence algorithm. However, numerical errors persisted when dealing with thin, strongly absorbing shells. Kaiser and Schweiger \cite{kaiser93} proposed a numerically stable algorithm for calculating Mie scattering coefficients of coated spheres, but it could not be easily extended to arbitrary multilayered spheres. Building on the work of Wu and Wang \cite{wu91}, Yang \cite{yang03} proposed an improved recursive algorithm for stable calculations of Mie scattering coefficients in complex cases, such as multilayered and strongly absorbing media. In the field of electromagnetics, Chew \cite{chew95} developed a layered medium framework suitable for spherical cases, which resulted in concise recurrence formulas. Recently, Yuan et al. \cite{yuan23} introduced these ideas into the spherical layered medium framework, yielding a set of numerically stable calculation schemes. In mathematical physics, Moroz \cite{moroz05} introduced the transfer matrix method to address the electromagnetic wave scattering problem in multilayered spheres. This approach, referred to as the Recursive Transfer Matrix Method (RTMM) by Moroz, is denoted here as the Recursive Transfer Matrix Algorithm (RTMA) to emphasize its algorithmic implementation. The transfer matrix method is an important and highly efficient technique for solving general scattering problems, although it still has some limitations in terms of applicability \cite{pena09, rasskazov20}.

To expand its applicability, this paper discusses an improved transfer matrix algorithm for the efficient and stable calculation of electromagnetic scattering and radiation by multilayered spheres. Our derivation of the spherical transfer matrix differs slightly from Moroz's approach, as we use Debye potentials to simplify the expressions, providing a new recursive format and a numerically stable algorithm that avoids overflow issues. Section 2 introduces the transfer matrix method for electromagnetic wave scattering by a multilayered sphere and presents new recursive formulas. Section 3 demonstrates a stable calculation scheme for the recursive formulas obtained through logarithmic derivatives or ratios of Bessel functions, with the final algorithm effectively avoiding round-off errors and numerical overflow issues caused by the imaginary part of large complex variables. Section 4 discusses three numerical test cases, showcasing the stability and applicability of the proposed algorithm. Section 5 concludes with a summary of the algorithm and results.

\section{Theory}
In spherical coordinates, the electromagnetic fields can be decomposed into transverse electric (TE) waves and transverse magnetic (TM) waves. For TE waves, $E^{\rm TE} = \nabla \times (\mathbf{r} \, \pi_{m}).$ For TM waves, $H^{\rm TM} = \nabla \times (\mathbf{r} \, \pi_{e}).$ The two scalar functions $\pi_{m}$ and $\pi_{e}$ are the so-called Debye potentials, which satisfy the scalar Helmholtz equation.
\begin{eqnarray}
(\nabla^2 + k^2)\pi_{e/m} = 0
\end{eqnarray}
with the wavenumber $k = \sqrt{\omega^2 \epsilon(r)\mu(r)}$, $\omega$ is the vacuum wavelength, and $\epsilon$ and $\mu$ represent the permittivity and permeability of the medium, respectively.

\begin{figure}
\centering
\begin{subfigure}{0.45\textwidth}
    \includegraphics[width=\textwidth]{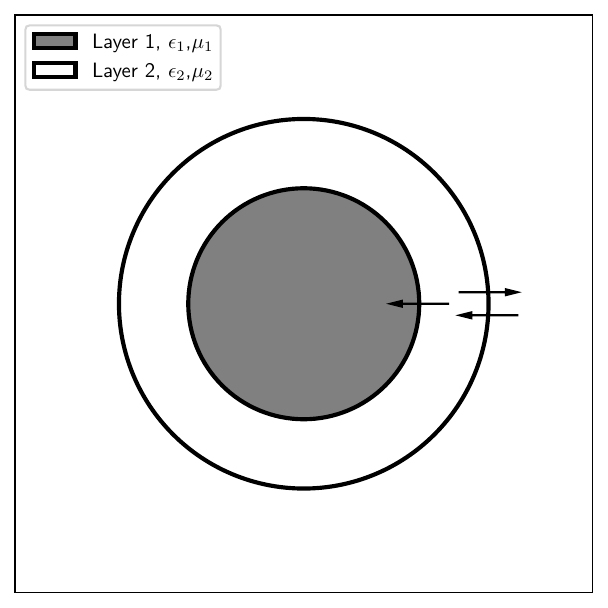}
    \caption{Standing wave scenario}
    \label{fig:fig2a}
\end{subfigure}
\hfill
\begin{subfigure}{0.45\textwidth}
    \includegraphics[width=\textwidth]{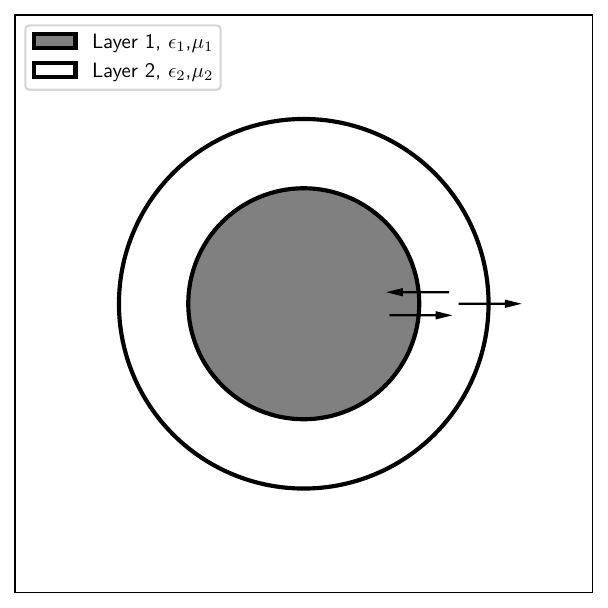}
    \caption{Outgoing wave scenario}
    \label{fig:fig2b}
\end{subfigure}
\caption{Schematic of an $n$-layer sphere embedded in an infinite dielectric medium. The outer radius, permittivity, and permeability of each layer are $r_i$, $\epsilon_i$, and $\mu_i$, respectively, with $i = 1, 2, ..., n$. The surrounding medium has vacuum permittivity $\epsilon_{n+1} = \epsilon_0$.}
\label{fig:sphere}
\end{figure}

Due to spherical symmetry of multilayered spheres (see Figure \ref{fig:sphere}), Debye potentials can be expanded as a sum of discrete modes $R_l(r) P^{m}_l(\cos\theta)\exp(im\phi)$. $\epsilon(r)$ and $ \mu(r)$ are constant within each spherical layer, the general form of the radial function $R_l(r)$ within each layer is a linear combination of spherical Bessel functions:
\begin{equation}\label{eq:3}
R_l(r)\big|_i = A_i j_l(k_i r) + B_i h^{(1)}_l(k_i r)
\end{equation}
where $j_l(z)$ and $h^{(1)}_l(z)$ are the spherical Bessel functions of the first and third kind, respectively. The expansion system of the scattered electromagnetic field, however, is composed of the corresponding Riccati-Bessel functions:
\begin{eqnarray}
\psi_l(z) = z j_l(z),\qquad \xi_l(z) = z h^{(1)}_l(z)\nonumber
\end{eqnarray}

The coefficient vectors $c_i^T = (A_i, B_i)^T$ for adjacent spherical layers are related by the boundary conditions at the interface, where the tangential components of the electromagnetic fields are continuous. Then, the boundary conditions at $r = r_i$ lead to 
\begin{align}
\Lambda^{\sigma}_{i+1}\Phi_{i+1} c_{i+1} = \Lambda^{\sigma}_{i}\Phi_{i} c_{i}
\end{align}
where $\sigma = m/e$ denotes polarization, the fundamental matrix $\Phi_i$ is given by
\begin{align}
\Phi_i(r)=
\begin{pmatrix}
\psi_l (k_{i}r) & \xi_l (k_{i}r)\\
\psi'_l (k_{i}r) &\xi'_l (k_{i}r)
\end{pmatrix}
\end{align}

Its inverse matrix is given by
\begin{align}
\Phi_i^{-1}(r)=\frac{1}{w_l(k_{i}r)}
\begin{pmatrix}
\xi'_l (k_{i}r) & -\xi_l (k_{i}r)\\
-\psi'_l (k_{i}r) &\psi_l (k_{i}r)
\end{pmatrix}
\end{align}
with the Wronskian of $\psi_l(k_{i}r) $ and $\xi_l(k_{i}r) $,
\begin{equation}
w_l(k_{i}r) = \psi_l (k_{i}r) \xi'_l (k_{i}r)  -\xi_l (k_{i}r) \psi'_l (k_{i}r)=\mathrm{i}
\end{equation}
with $\mathrm{i} = \sqrt{-1}$. Furthermore, $\Lambda^{\sigma}_{i}$  is a diagonal matrix, and its specific form depends on the convention. For transverse magnetic (TM) polarization,
\begin{align}
\Lambda^{m}_{i} =
\begin{pmatrix}
 \frac{\epsilon_{i}}{k_{i}}& 0\\
0 & 1
\end{pmatrix}
\end{align}
The transverse electric (TE) $\Lambda^{e}_{i}$ can be obtained by the duality principle.

The matrix $\Lambda^{\sigma}_{i}$ acts as a scaling factor. For the sake of simplicity in the derivation, we can incorporate it into the fundamental matrix, such as $\frac{\epsilon_i}{k_i} \psi_l(k_i r_i) \to \tilde{\psi}_l(k_i r_i)$. Without affecting understanding, we omit the superscript $\tilde{\cdot}$ and eventually restore the final expression. Multiplying both sides of the equation by $\Phi^{-1}_{i+1}$, we get
\begin{align}
\begin{pmatrix}
A_{i+1}\\
B_{i+1}
\end{pmatrix}
=
M_{i+1, i} 
\begin{pmatrix}
A_i\\
B_i
\end{pmatrix}
\label{eq:8}\end{align}
where the matrix $M_{i+1, i}$ connects the coefficients of adjacent layers and is called the transfer matrix.
\begin{eqnarray}
M_{i+1, i}&=&\Phi_{i+1}^{-1} \Phi_i\nonumber\\
&=&\frac{1}{w_l(z^+_i ) }
\begin{pmatrix}
 \psi_l (z_i)  \xi'_l (z^+_i )   - \psi'_l (z_i) \xi_l (z^+_i ) & \xi'_l (z^+_i )\xi_l (z_i) -\xi_l (z^+_i )\xi'_l (z_i)\nonumber\\
-\psi_l (z_i) \psi'_l (z^+_i )  + \psi'_l (z_i) \psi_l (z^+_i ) & -\psi'_l (k_{i+1}r_i)\xi_l (z_i) + \psi_l (z^+_i )\xi'_l (z_i)
\end{pmatrix}    
\end{eqnarray}
with $z_i = k_i r_i$, $z^+_i =k_{i+1}r_i$. And $w_l(z^+_i ) $ is the the Wronskian of $\tilde{\psi_l}(z^+_i) $ and $\tilde{\xi_l}(z^+_i) $, which equals $\mathrm{i} \epsilon_{i+1}/k_{i+1}$ for the TM mode.

For an $n$-layer sphere, we have the following formula
\begin{align}
\begin{pmatrix}
A_{n+1}\\
B_{n+1}
\end{pmatrix}
=
M_{n+1, n} M_{n, n-1}...M_{3,2}M_{2,1}
\begin{pmatrix}
A_1\\
B_1
\end{pmatrix} = M
\begin{pmatrix}
A_1\\
B_1
\end{pmatrix}
\end{align}
Since the third kind of Bessel functions are singular at the origin, we set $A_1 = 1$ and $B_1 = 0$. Therefore, the coefficients in the outermost medium are directly given by the matrix elements of $2 \times 2$ matrix $M$, i.e., $A_{n+1} = M^{11}$ and $B_{n+1} = M^{21}$. The total scattering coefficients for the $n$-layer sphere are given by the following formula:
\begin{equation}
\mathcal{T}_{\sigma, l} = \frac{B_{n+1}}{A_{n+1}} =\frac{M^{11} }{M^{21}}
\end{equation}

Due to the spherical symmetry, the T matrix of the multilayered sphere is diagonal, and its diagonal elements correspond exactly to the Mie scattering coefficients.
\[
\mathcal{T}_{\mu\mu'} = \mathcal{T}_{\sigma, l} \delta_{\sigma\sigma'} \delta_{l,l'}\delta_{mm'}
\]
$\mu = \{\sigma, l, m\}$ is the index of the supermatrix. By summing over all TM and TE polarizations and channel contributions, the scattering and extinction efficiencies are determined by:
\begin{equation}
Q_{sca} = \frac{2}{x^2} \text{Tr} \Big(\mathcal{T}^{\dagger} \mathcal{T} \Big)
\end{equation}
\begin{equation}
Q_{ext} = -\frac{2}{x^2} \text{Tr}\Big(\frac{\mathcal{T}^{\dagger} + \mathcal{T}}{2}  \Big)
\end{equation}
Here,  $x = 2 \pi R/ \lambda$ is the size parameter, $R$ is the radius of the outermost layer of the sphere, and $\lambda$ is the wavelength in vacuum. The absorption efficiency is given by $Q_{abs} = Q_{ext}- Q_{sca}$.  The above expression is also applicable to irregularly shaped objects, such as those discussed in our previous work \cite{zhang16} on scattering from randomly shaped geometries, and spheres embedded in an absorbing medium \cite{zhang24}. The energy emissivity of the multilayered sphere is given by the following formula \cite{kruger12, kattawar70}:
\begin{eqnarray}
H_s &=&  \frac{8\pi^2 R^2 h c^2}{\lambda^5}   \int^{\infty}_0 d\lambda \frac{1}{\exp{\frac{h c}{k_BT\lambda}}-1}Q_{abs}
\end{eqnarray}

Here, $k_B$ and $h$ are the Boltzmann and Planck constants, respectively, and $c$ is the speed of light in vacuum.

\section{Algorithm}
In this section, we present a recursive algorithm based on the transfer matrix method, referred to as the modified Recursive Transfer Matrix Algorithm (mRTMA). It is important to note that while the RTMA involves the combined use of the transfer matrix, it does not align with the definition of a recursive algorithm as commonly understood in the field of scientific computing.
 
The scattering coefficient for each layer can be given by the following equation:
\[
\mathcal{T}^{i}_{\sigma,l}  = \frac{B_i}{A_i}
\]
 Additionally, we obtain a set of recursive equations:
\begin{eqnarray}
\mathcal{T}^{i+1}_{\sigma,l} = \frac{M^{21}_{i+1, i} + M^{22}_{i+1, i} \mathcal{T}^{i}_{\sigma,l}  }{M^{11}_{i+1, i} + M^{12}_{i+1, i} \mathcal{T}^{i}_{\sigma,l}  }
\end{eqnarray}
with
\[
\mathcal{T}^{0}_{\sigma,l}  = 0.
\]
Substituting the elements of the transfer matrix (eq.\ref{eq:8}) into the above expression, we have:
\begin{eqnarray}
\mathcal{T}^{i+1}_{\sigma,l} = \frac{ \psi'_l (z_i) \psi_l (z^+_i) -\psi_l (z_i) \psi'_l (z^+_i) + (\psi_l (z^+_i)\xi'_l (z_i)-\psi'_l (z^+_i)\xi_l (z_i)  ) \mathcal{T}^{i}_{\sigma,l} }{ \psi_l (z_i)  \xi'_l (z^+_i)   - \psi'_l (z_i) \xi_l (z^+_i) + (\xi_l (z_i)\xi'_l (z^+_i)  -\xi'_l (z_i)\xi_l (z^+_i)) \mathcal{T}^{i}_{\sigma,l}  }
\end{eqnarray}
When the refractive index of the sphere is real, the arguments of the Bessel function and its derivative are also real, making it easy to obtain robust recurrence formulas for their evaluation. However, when the refractive index is complex, the calculation of the Bessel function becomes unstable as it enters the exponential domain and goes out of bounds.
To avoid this ill-conditioning, the calculation formula will involve only the ratios and logarithmic derivatives of the Bessel functions. 

The logarithmic derivatives are 
\begin{eqnarray}
D_l(z) = \frac{\psi'_l(z)}{\psi_l(z)},\qquad F_l(z) =\frac{\xi'_l(z)}{\xi_n(z)}
\end{eqnarray}

For the case of a multilayer sphere, $\mathcal{T}^{i}_{\sigma, l}$ may still encounter numerical overflow issue. To avoid this issue, we define a rescaled scattering coefficient $\tilde{\mathcal{T}}^{i}_{\sigma,l}$ and a combined ratio $Q^i_{l}$ \cite{yang03}
\[
\tilde{\mathcal{T}}^{i}_{\sigma,l}   = \frac{\xi_l (z^+_{i-1})}{\psi_l (z^+_{i-1})} \mathcal{T}^{i}_{\sigma,l},\qquad Q^i_{l} = \frac{\psi_l (z^+_{i-1}) \xi_l (z_i)}{\psi_l (z_i) \xi_l (z^+_{i-1}) }
\]

Then, the recurrence formula becomes
\begin{eqnarray}
\tilde{\mathcal{T}}^{i+1}_{\sigma,l}  = -\frac{\big(D_l (z^+_i)  -  D_l(z_i)\big) +  Q^i_{l}  \big(D_l (z^+_i)- F_l (z_i)\big) \tilde{\mathcal{T}}^{i}_{\sigma,l}  }{ \big(F_l (z^+_i)   - D_l (z_i)\big) + Q^i_{l} \big( F_l  (z^+_i)-F_l  (z_i) \big) \tilde{\mathcal{T}}^{i}_{\sigma,l}   }
\end{eqnarray}
with 
\[
\tilde{\mathcal{T}}^{0}_{\sigma,l}  = 0
\]
Consequently, we can recover the formula for TM wave,
\begin{eqnarray}
\tilde{\mathcal{T}}^{i+1}_{m,l}  = -\frac{\big(\sqrt{\epsilon_{i}\mu_{i+1}}D_l (z^+_i)  - \sqrt{\epsilon_{i+1}\mu_{i}} D_l(z_i)\big) +  Q^i_{l}  \big(\sqrt{\epsilon_{i}\mu_{i+1}}D_l (z^+_i)- \sqrt{\epsilon_{i+1}\mu_{i}}F_l (z_i)\big) \tilde{\mathcal{T}}^{i}_{m,l}  }{ \big(\sqrt{\epsilon_{i}\mu_{i+1}}F_l (z^+_i)   - \sqrt{\epsilon_{i+1}\mu_{i}}D_l (z_i)\big) + Q^i_{l} \big( \sqrt{\epsilon_{i}\mu_{i+1}}F_l  (z^+_i)-\sqrt{\epsilon_{i+1}\mu_{i}}F_l  (z_i) \big) \tilde{\mathcal{T}}^{i}_{m,l}  } \nonumber
\end{eqnarray}
By duality, then
\begin{eqnarray}
\tilde{\mathcal{T}}^{i+1}_{e,l}  = -\frac{\big(\sqrt{\epsilon_{i+1}\mu_{i}}D_l (z^+_i)  - \sqrt{\epsilon_{i}\mu_{i+1}} D_l(z_i)\big) +  Q^i_{l}  \big(\sqrt{\epsilon_{i+1}\mu_{i}}D_l (z^+_i)- \sqrt{\epsilon_{i}\mu_{i+1}}F_l (z_i)\big) \tilde{\mathcal{T}}^{i}_{e,l}  }{ \big(\sqrt{\epsilon_{i+1}\mu_{i}}F_l (z^+_i)   - \sqrt{\epsilon_{i}\mu_{i+1}}D_l (z_i)\big) + Q^i_{l} \big( \sqrt{\epsilon_{i+1}\mu_{i}}F_l  (z^+_i)-\sqrt{\epsilon_{i}\mu_{i+1}}F_l  (z_i) \big) \tilde{\mathcal{T}}^{i}_{e,l}  } \nonumber
\end{eqnarray}
In addition to the above standing wave case, we can also consider the outgoing waves case \cite{chew95}. Similarly, we define
\[
\breve{\mathcal{T}}^{i}_{\sigma,l}   = \frac{\psi_l (z_{i})}{\xi_l (z_{i})} \mathcal{T}^{i}_{\sigma,l}
\]
Then, it is easy to obtain the rescaled Mie scattering coefficient for the outgoing wave case,
\begin{eqnarray}
\breve{\mathcal{T}}^{i}_{\sigma,l}  &=& -\frac{  \big( F_l (z^+_i) - F_l (z_i)  \big) + Q^{i+1}_{l}   \big( D_l (z^+_i)  - D_l(z_i)  \big) \breve{\mathcal{T}}^{i+1}_{\sigma,l}   }{ 
  \big(F_l (z^+_i) - D_l(z_i) \big) + Q^{i+1}_{l}  \big( D_l (z^+_i)-D_l (z_i)  \big) \breve{\mathcal{T}}^{i+1}_{\sigma,l}  }
\end{eqnarray}
Consequently, we can recover the formula for TM wave,
\begin{eqnarray}
\breve{\mathcal{T}}^{i}_{m,l}  &=& -\frac{  \big( \sqrt{\epsilon_{i}\mu_{i+1}} F_l (z^+_i) - \sqrt{\epsilon_{i+1}\mu_{i}} F_l (z_i)  \big) + Q^{i+1}_{l}    \big( \sqrt{\epsilon_{i}\mu_{i+1}}D_l (z^+_i)  - \sqrt{\epsilon_{i+1}\mu_{i}} D_l(z_i) \big) \breve{\mathcal{T}}^{i+1}_{m,l}   }{ 
  \big(\sqrt{\epsilon_{i}\mu_{i+1}}F_l (z^+_i) -\sqrt{\epsilon_{i+1}\mu_{i}}  D_l(z_i) \big) + Q^{i+1}_{l}  \big( \sqrt{\epsilon_{i}\mu_{i+1}}D_l (z^+_i)- \sqrt{\epsilon_{i+1}\mu_{i}} D_l (z_i)  \big) \breve{\mathcal{T}}^{i+1}_{m,l}  } \nonumber
\end{eqnarray}
By duality, then
\begin{eqnarray}
\breve{\mathcal{T}}^{i}_{e,l}  &=& -\frac{ \big( \sqrt{\epsilon_{i+1}\mu_{i}} F_l (z^+_i) - \sqrt{\epsilon_{i}\mu_{i+1}} F_l (z_i)  \big) + Q^{i+1}_{l}    \big( \sqrt{\epsilon_{i+1}\mu_{i}}D_l (z^+_i)  - \sqrt{\epsilon_{i}\mu_{i+1}} D_l(z_i)  \big) \breve{\mathcal{T}}^{i+1}_{e,l}   }{ 
  \big(\sqrt{\epsilon_{i+1}\mu_{i}}F_l (z^+_i) -\sqrt{\epsilon_{i}\mu_{i+1}}  D_l(z_i)  \big) + Q^{i+1}_{l}  \big( \sqrt{\epsilon_{i+1}\mu_{i}}D_l (z^+_i)- \sqrt{\epsilon_{i}\mu_{i+1}} D_l (z_i)  \big) \breve{\mathcal{T}}^{i+1}_{e,l}  } \nonumber
\end{eqnarray}

The downward recursion for $D_l$ is numerically stable. Therefore, we use the following recursive formula for computation \cite{kattawar67}:
\begin{eqnarray}
D_{l-1}(z) &=& \frac{l}{z} - \Big(\frac{l}{z} + D_{l}(z)\Big)^{-1}
\end{eqnarray}
The upward recursion for $F_l$ is numerically stable. Therefore, we use the following recursive formula for computation:
\begin{eqnarray}
F_l(z) &=&  \Big(\frac{l}{z} - F_{l-1}(z)\Big)^{-1} - \frac{l}{z} 
\end{eqnarray}

Using logarithmic derivatives, Toon and Ackerman \cite{toon81} provided a stable recursive algorithm for the ratio of Bessel functions $\frac{ \psi_l (z^+_{i-1}) }{\psi_l (z_{i}) }$. Similarly, the ratio $ \frac{\xi_l (z_{i}) }{\xi_l (z^+_{i-1}) }$ can be computed using a similar approach. Yang \cite{yang03} combined these two ratios into a single calculation.
\begin{eqnarray}
Q^i_{l} = \frac{D_l(z_i) + l/z_i}{D_l(z^+_{i-1}) + l/z^+_{i-1}}  \frac{F_l(z^+_{i-1}) + l/z^+_{i-1}} {F_l(z_i) + l/z_i}Q^i_{l-1}  
\end{eqnarray}

with initial values,
\begin{eqnarray}
Q^i_{0}= e^{-2(\text{Im}z^+_{i-1} - \text{Im}z_{i})}\frac{ e^{-2\text{Re}z^+_{i-1}} - e^{-2\text{Im}z^+_{i-1}}}{  e^{-2\text{Re}z_{i}}- e^{-2\text{Im}z_{i}}} 
\end{eqnarray}

The main contribution of this paper is the improvement of the transfer matrix method, making it suitable for recursive computation, which can be applied to cases with an ultra-high number of layers and strongly absorbing media. Compared to the works of Wu \cite{wu91} and Yang \cite{yang03}, our recursive formulas are more concise and allow for the sequential computation of expansion coefficients for each spherical layer, making the algorithm more efficient. Additionally, we have derived recursive formulas applicable to the outgoing wave scenario.

We next address potential issues that may arise during the implementation of our algorithm and propose corresponding solutions. For real arguments $z$ (layers with purely real material constants), $\psi_l$ is known to exhibit real zeros, leading to singularities in $D_l$ as noted by one of the reviewers. To handle this, a hybrid recursive scheme combining Eq. 15 and Eq. 17 can be employed. Specifically, for spherical layer that encounter the singularity of $D_l$, Eq. 15 can be applied for computation. This approach represents another advantage of the mRTMA. 
For complex arguments $z$, $\xi_l$ diverges to infinity while $\psi_l$ approaches zero as $l$ goes to infinity. However, practical computations involve a truncation order, $l_{max}$, approximately $|z| + p|z|^{1/3} + 1$, where $p$ is set to be about 4. Within the truncation range ($l < l_{max}$), the ratio $|\xi_l /\psi_l |$ increases monotonically. Thus, ensuring the boundedness of $|\xi_{l_{max}} /\psi_{l_{max}} |$ suffices for stability. Empirically, it can be observed that $\xi_l$ and $\psi_l$ are proportional to $e^{-\text{Im} z}$ and $e^{\text{Im} z}$, respectively, causing $|\xi_l/\psi_l|$ decay exponentially with $2\text{Im} z$. At the truncation order, $l_{max}$, it is straightforward to verify that $|\xi_l/\psi_l| < 1$.

\section{Results}
The following section presents three test cases using the algorithm proposed in this paper. The first test case examines the scattering problem of a double-layered sphere with varying radii, where the refractive indices of the inner and outer layers are specified and compared with the RTMA. The second test case involves scattering by a multilayered sphere with the same radius, where the refractive indices of each shell are randomly generated, and the extinction cross-section is calculated for spheres with various numbers of layers. The third test case addresses the thermal radiation problem of a multilayered sphere, calculating the emissivity of a double-layer sphere.
\begin{figure}
\centering
\begin{subfigure}{0.7\textwidth}
    \includegraphics[width=\textwidth]{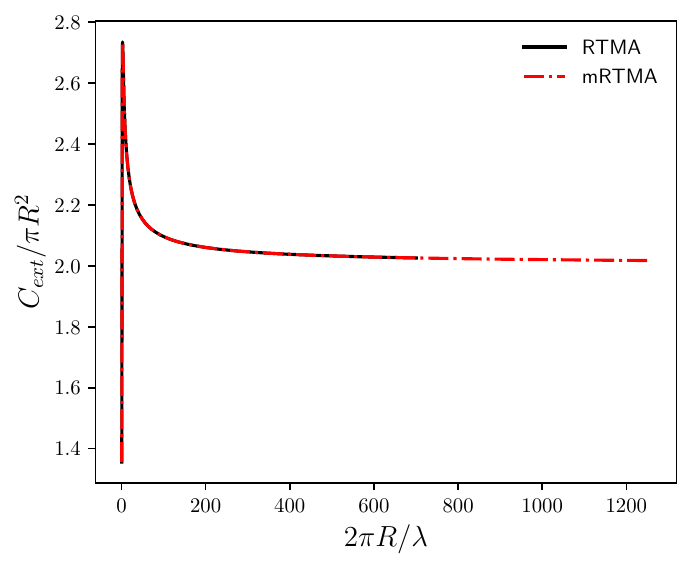}
    \caption{}
    \label{fig:fig2a}
\end{subfigure}
\hfill
\begin{subfigure}{0.7\textwidth}
    \includegraphics[width=\textwidth]{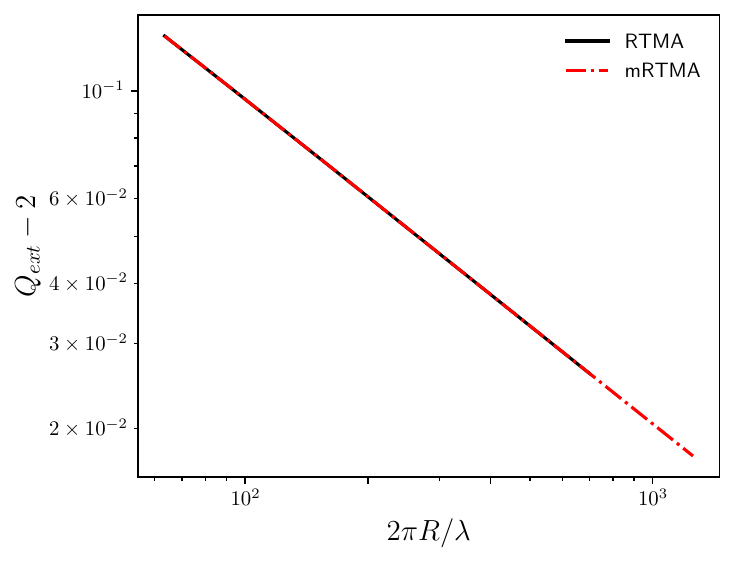}
    \caption{}
    \label{fig:fig2b}
\end{subfigure}
\caption{(a) Plot of the extinction efficiency $Q_{ext}$ for the coated sphere as a function of the outermost layer's size parameter. (b) Loglog plot of $Q_{ext}$ - 2 of the coated sphere as a function of the outermost layer's size parameter.  The plot compares the results with those obtained using RTMA. The refractive indices of the core and shell are  $m_1=1.33+0.0\mathrm{i}$ and $m_2=1.33+\mathrm{i}$, respectively.}
\label{fig:fig2}
\end{figure}
In the 1st test, a coated sphere is considered, with the refractive indices of the core and shell layers set to $m_1 = 1.33 + 0.0\mathrm{i}$ and $m_2 = 1.33 + \mathrm{i}$, respectively, and the core radius accounting for 50\% of the total radius. The extinction efficiency $Q_{ext}$ is calculated for size parameters $x \in [1.0, 1200.0]$. The RTMA experiences numerical overflow for $x > 700.0$, while the mRTMA remains stable, as shown in Figure \ref{fig:fig2}.

\begin{figure}
\centering
\begin{subfigure}{0.7\textwidth}
    \includegraphics[width=\textwidth]{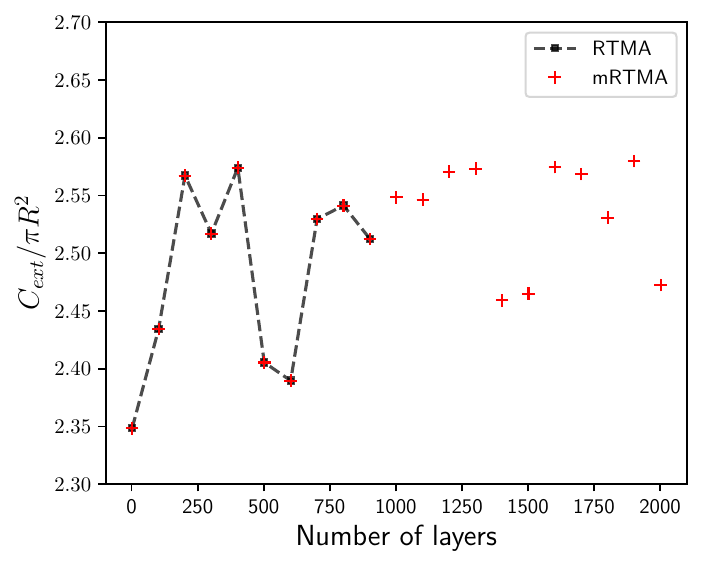}
    \caption{}
    \label{fig:fig3a}
\end{subfigure}
\hfill
\begin{subfigure}{0.7\textwidth}
    \includegraphics[width=\textwidth]{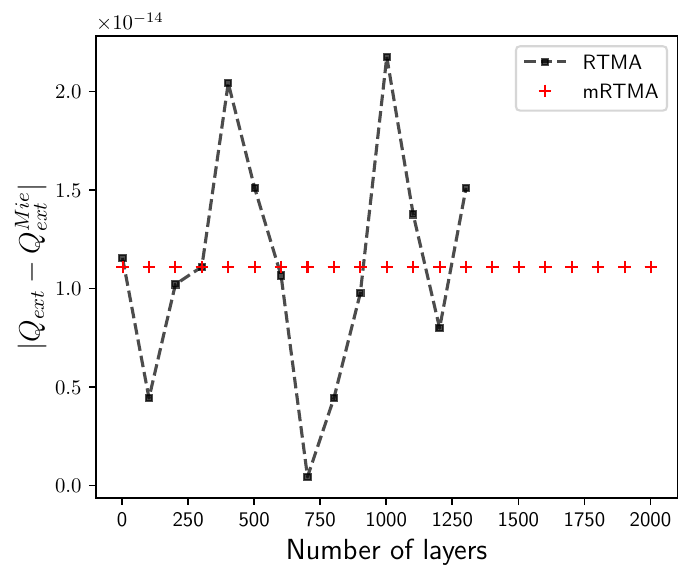}
    \caption{}
    \label{fig:fig3b}
\end{subfigure}
\caption{(a) The extinction efficiency $Q_{\rm ext}$ of the multilayer sphere as a function of the number of layers. The refractive index of each layer is randomly generated. (b) Absolute difference between $Q_{\rm ext}$ and $Q_{\rm ext}^{\rm Mie}$, where $Q_{\rm ext}^{\rm Mie}$ is the extinction efficiency for a homogeneous sphere, computed using the Mie algorithm \cite{mishchenko02}. The refractive index is fixed at $1.33 + \mathrm{i}$. The size parameter of the sphere in (a) and (b) is $4\pi$.}
\label{fig:fig3}
\end{figure}
In the 2nd test, we examined the scattering of a multilayered sphere with randomly generated refractive indices, with the total number of layers ranging from 2 to 2002. The refractive indices are generated as $n = n' + \mathrm{i} n''$, with $n' \sim \mathcal{U}_{[1, 2]}$ and $\log_{10} n'' \sim \mathcal{U}_{[-3, 1]}$, where $\mathcal{U}$ denotes a uniform distribution. Additionally, we also tested the case of a multilayered sphere where all layers have the same refractive index of $1.33+ \mathrm{i}$, and compared the results with the Mie scattering results for a homogeneous sphere. As shown in Figure \ref{fig:multi}, the mRTMA demonstrates significant improvement over the RTMA.

\begin{figure}
 \centering
\includegraphics[width=0.7\linewidth]{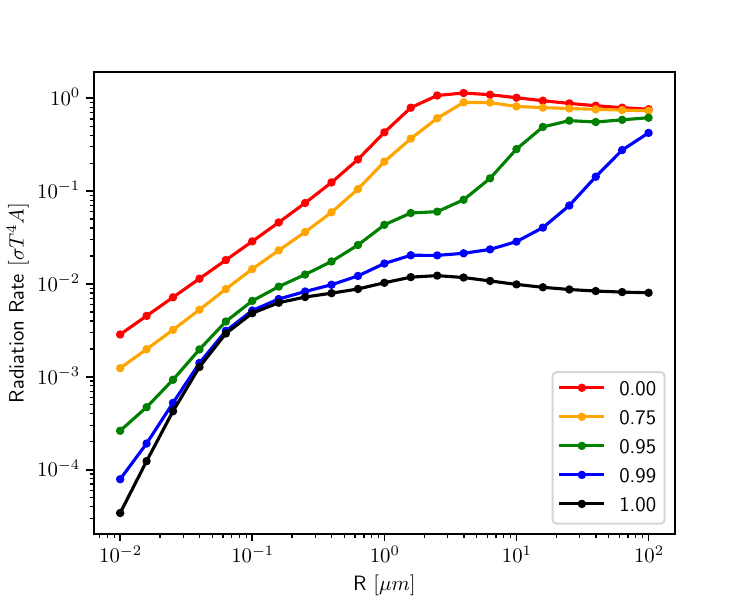}
\caption{The emitted energy of a SiC+gold coated sphere at a temperature of 300K, as a function of the outer radius R, is normalized with respect to the Stefan-Boltzmann result. Different curves correspond to different ratios of inner to outer layer radii. For SiC and gold, the optical constants' data are from \cite{laor93} and \cite{johnson72} respectively.}
  \label{fig:fig4}
\end{figure}
In the 3rd test, the mRTMA is used to calculate the electromagnetic radiation of a double-layer sphere with varying radii, with the sphere maintained at a temperature of 300K in a cold environment. Figure \ref{fig:fig4} shows the radiation characteristics of the double-layer sphere, using SiC and gold as example materials, representing a dielectric and a conductor, respectively.

\begin{figure}
 \centering
\includegraphics[width=0.7\linewidth]{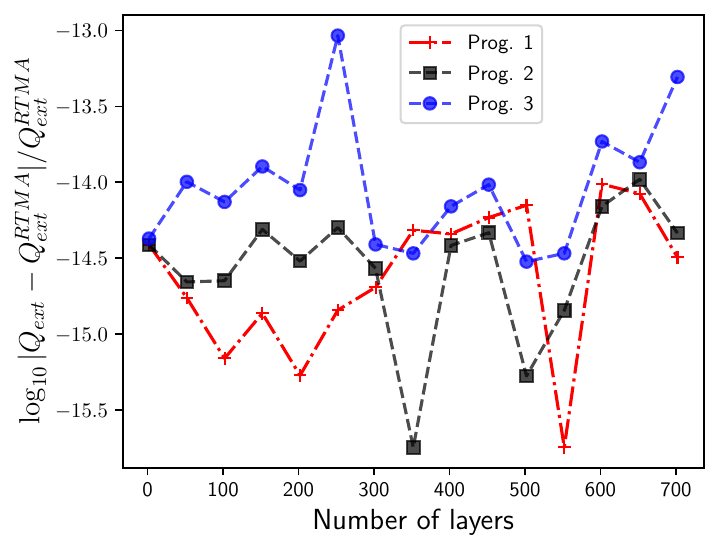}
\caption{Performance validation. Prog. 1: Our implementation of the mRTMA. Prog. 2: Our implementation of Yang's recursive algorithm. Prog. 3: Scattnlay's implementation of Yang's recursive algorithm. Size parameter of the sphere is $4\pi$. }
  \label{fig:fig5}
\end{figure}
In the 4th test, we validate the precision of mRTMA by comparing it with RTMA and Yang's recursive algorithm. The multilayered sphere is simulated using randomly generated refractive indices, with the total number of layers ranging from 2 to 702, increasing in steps of 50 layers. The refractive indices are generated as $n = n' + \mathrm{i} n''$, with $n' \sim \mathcal{U}_{[1, 2]}$ and $\log_{10} n'' \sim \mathcal{U}_{[-3, 1]}$, where $\mathcal{U}$ denotes a uniform distribution. As shown in Figure \ref{fig:fig5}, the RTMA, mRTMA, and Yang's recursive algorithm exhibit excellent agreement with high precision in this case, demonstrating the validity of the mRTMA.

\section{Conclusion}

This paper aimed to develop a new, eﬀicient algorithm for electromagnetic radiation and scattering by a multilayered sphere based on Lorentz-Mie theory, with a particular emphasis on addressing numerical stability issues. We examined the evolution of various improved algorithms, summarizing and comparing recursive methods and transfer matrix algorithms to identify their limitations and potential challenges. These challenges are especially pronounced in scenarios involving thin shells and strongly absorbing layers.

Building on earlier works, we proposed an efficient recursive transfer matrix algorithm (mRTMA). The primary contribution of mRTMA lies in the development of new recursive formulas based on the Transfer Matrix Method (RTMA). These formulas effectively address the numerical overflow issues encountered in RTMA and enable the algorithm to handle multilayer spheres embedded in absorptive host media. Like RTMA, mRTMA can simultaneously compute the expansion coefficients for spherical layers.

Moreover, by adopting a hybrid recursive strategy, mRTMA resolved the singularity problem associated with logarithmic derivatives in Yang’s algorithm. In addition, unlike both RTMA and Yang’s algorithm, mRTMA introduced recursive formulas specifically designed for the outgoing wave scenario, where the incident wave originates from within the sphere. These formulas broaden the applicability of the method to specific scenarios, such as spherical cavities and spherical antennas.

Finally, we demonstrated that mRTMA achieves high accuracy across a broad range of size parameters and significantly extends the applicability of classical transfer matrix methods. Future work can focus on optimizing these numerical methods to further enhance their stability and efficiency, providing more effective tools for solving complex electromagnetic scattering problems.

%






%
%
%
\bibliographystyle{unsrt}  
\bibliography{reference} 
\end{document}